\documentclass[11pt]{article}
\usepackage{verbatim}
\usepackage[margin=1in]{geometry}
\usepackage{amssymb,amsmath}
\usepackage{graphicx}
\usepackage{setspace}
\usepackage{sidecap}
\usepackage{amsthm}
\usepackage{enumerate}
\usepackage{multirow}
\usepackage[table]{xcolor}
\usepackage{booktabs}
\usepackage{makecell}

\usepackage[colorlinks = true,
            linkcolor = blue,
            urlcolor  = blue,
            citecolor = blue,
            anchorcolor = blue]{hyperref}

\usepackage[title]{appendix}

\usepackage{cleveref}  
\usepackage{cite}

\crefname{appendix}{Appendix}{Appendices}
\Crefname{appendix}{Appendix}{Appendices}

\usepackage{longtable}
\usepackage{siunitx}

\usepackage[font={small,it}]{caption}

\usepackage{tikz}
\usetikzlibrary{shapes,arrows}

\tikzstyle{decision} = [diamond, draw, text width=4.5em, text badly centered, node distance=3cm, inner sep=0pt]
\tikzstyle{block} = [rectangle, draw, text width=5em, text centered, rounded corners, minimum height=4em]

\begin{document}

\title{Stochasticity and Practical Identifiability in Epidemic Models: A Monte Carlo Perspective}

\author{Chiara Mattamira$^1$, Olivia Prosper Feldman$^2$
\\\footnotesize$^{1,}$ $^2$ Department of Mathematics, University of Tennessee, Knoxville, TN, USA,\\ }
\date{}
\maketitle



\begin{abstract}
Assessing the practical identifiability of epidemic models is essential for determining whether parameters can be meaningfully estimated from observed data. Monte Carlo (MC) methods provide an accessible and intuitive framework; however, their standard implementation—perturbing deterministic trajectories with independent Gaussian noise—rests on assumptions poorly suited to epidemic processes, which are inherently stochastic, temporally correlated, and highly variable, especially in small populations or under slow transmission. In this study, we investigate the structure of stochastic variability in the classic Susceptible–Infected–Recovered (SIR) model across a range of epidemiological regimes, and assess whether it can be represented within the independent Gaussian noise framework. We show that continuous-time Markov chain (CTMC) trajectories consistently exhibit super-Poissonian variability and strong temporal dependence. Through coverage analysis, we further demonstrate that independent Gaussian noise systematically underestimates the variability of the underlying stochastic process, leading to overly optimistic conclusions about parameter identifiability. In addition, we propose a hybrid simulation approach that introduces time- and amplitude-dependent variability into deterministic ODE trajectories, preserving computational efficiency while capturing key features of epidemic stochasticity. Our findings highlight the limitations of the standard MC algorithm and provide a pathway for incorporating more realistic noise structures into epidemic inference.
\\[1ex]
\end{abstract}

\newpage
\section{Introduction}

The use of mathematical and statistical models to provide outbreak predictions and inform response strategies has gained prominence in recent years, especially following 
the 2019 Measles outbreak in the Democratic Republic of Congo \cite{davis2021measles}, the global spread of COVID-19 \cite{miranda2021covid,olabode2021covid, zhao2020covid}, and more recently, the 2022 Mpox outbreak \cite{srivastava2023global, ward2022transmission}, as researchers increasingly recognize the need for these advanced quantitative tools. These outbreak predictions often rely on estimates of \textit{model parameters} that capture key epidemiological quantities such as the basic reproduction number $(R_0)$ and the average length of the infectious and recovery periods. Consequently, it is crucial to evaluate the reliability of the parameter estimates, which begins with assessing a model's \textit{identifiability}. 

Given a model and experimental output, identifiability addresses the question of whether the model parameters can be \textit{uniquely} determined. This question is posed under two distinct scenarios that create the distinction between structural and practical identifiability. \textit{Structural} identifiability assumes perfectly observed data and an error-free model, asking whether, in principle, a unique parameter set could generate the observed behavior. In contrast, \textit{practical} identifiability considers real-world data subject to noise and model misspecification. Hence, assessing practical identifiability is especially important when making predictions and control strategy decisions driven by real-world data.
Several studies have demonstrated that a practically unidentifiable model-data combination can result in significantly divergent predictions \cite{eisenberg2013identifiability, kao2018practical, tuncer2018structural}. We include an illustrative example in \cref{appendix:control_strategy}, where practical unidentifiability prevents reliable evaluation of a given intervention strategy.

Several methods have been developed to assess practical identifiability with sensitivity-based techniques, profile likelihood analysis, Bayesian approaches, and Monte Carlo simulation-based methods being the most commonly used \cite{lam2022practical, miao2011identifiability, raue2009structural, wieland2021structural}. As highlighted in recent review papers \cite{lam2022practical, wieland2021structural}, each method trades off between computational cost, interpretability, and robustness. For example, sensitivity-based methods are computationally light, but their local linear approximations can give misleading results even for simple nonlinear models \cite{krausch2019monte, lam2022practical, wieland2021structural}. 
Profile likelihood methods have emerged as a popular choice due to their ability to detect both structural and practical non-identifiability and their usefulness in informing experimental design \cite{raue2009structural, wieland2021structural}. 
However, they rely on a correctly specified likelihood function and produce parameter-wise confidence intervals by fixing one parameter at a time while optimizing over the others, effectively slicing through the likelihood surface \cite{raue2009structural, robins2000comment}. This approach may miss identifiability issues that arise from interactions between parameters, as it does not reflect the full joint structure of the likelihood \cite{komorowski2011sensitivity}. In contrast, Bayesian methods  recover the full joint posterior distribution and naturally incorporate prior information \cite{van2014gentle}. Despite these advantages, their application often requires considerable computational resources and specialized expertise, which can limit their practical use in widespread identifiability assessments. Finally, Monte Carlo (MC) simulation benefits from being a relatively intuitive and simple way to quantify parameter uncertainty, but it can be computationally expensive. Moreover, as noted by Lam et al. \cite{lam2022practical}, a more fundamental limitation may lie in the fidelity of the artificially simulated noise to the uncertainty of real-world measurements. 

In this work, we focus on practical identifiability analysis of compartmental epidemic models using Monte Carlo simulation. These rely on generating synthetic datasets that reflect the uncertainty observed in real outbreaks. This is commonly done by solving the deterministic ODE model and adding independent Gaussian noise to the output 
\cite{chis2011structural, miao2011identifiability}. 
This simulation method is widely used for its conceptual simplicity and low computational cost \cite{liyanage2025structural, saucedo2024comparative, tuncer2018structural, yang2024parameter}. However, this approach does not distinguish between two fundamentally different sources of uncertainty: stochastic variation due to intrinsic epidemic dynamics and extrinsic measurement error. Independent Gaussian noise is typically used to model measurement error, such as instrument-level variability or reporting delays \cite{guo2011estimation}. This kind of noise is generally modeled as pointwise, uncorrelated, and arbitrarily scaled --- assumptions that may be reasonable when modeling white noise, but fail to capture important characteristics of real epidemic data, such as strong time dependence, structured fluctuations, and stochastic variation in peak timing and magnitude \cite{mamis2023stochastic}. Moreover, process-level stochasticity is frequently super-Poissonian, especially in slow transmission scenarios and, as we show in our coverage results, an unrealistically large noise level would be required to replicate the stochastic uncertainty.

To address these issues, we begin by analyzing the structure of stochastic deviations to understand the limitations of the assumptions behind independent Gaussian noise models. We then evaluate whether alternative distributions or empirical resampling can better capture stochastic variation, and show that the lack of temporal structure remains a key limitation. To overcome this, we introduce a hybrid method that modifies the deterministic ODE solution by applying time and amplitude warping, capturing key stochastic features without requiring full CTMC simulation. We also investigate how the magnitude of the applied time and amplitude shifts varies with $R_0$ and the total population size $N$, providing insight into how stochastic variability scales with epidemic parameters. We assess all simulation strategies by computing their coverage across a diverse set of parameter combinations representing realistic disease dynamics. We show that all independent noise models tested lead to poor coverage across all parameter combinations, while the hybrid approach achieves performance comparable to stochastic simulations.  Lastly, we compare the spread of best-fit parameter estimates across methods and highlight how underestimating stochastic uncertainty can lead to misleading identifiability results.

The remainder of our study is structured as follows. In \textsc{\cref{sec:setup}}, we define the identifiability framework, benchmark simulation methods, and parameter selection. In \textsc{\cref{sec:stochastic_analysis}}, we analyze the structure of stochastic deviations and evaluate alternative noise models. In \textsc{\cref{sec: hybrid_method}}, we introduce and assess our hybrid simulation approach. In \textsc{\cref{sec: coverage}} and \textsc{\cref{sec:identifiability}}, we benchmark all methods across a range of scenarios. We conclude with a discussion of key findings and implications in \textsc{Discussion}.

\section{Framework and Setup}\label{sec:setup}
\subsection{Monte Carlo Framework for Practical Identifiability}\label{sec:MC}
Given a model with state variable vector \textbf{x(t)} and parameter set $\mathbf{\hat p}$, which can be obtained from fitting to experimental data or drawn from literature values, the Monte Carlo algorithm for assessing practical identifiability proceeds as follows \cite{lam2022practical}:
\begin{enumerate}
    \item Solve the model numerically using the parameter set $\hat{\mathbf{p}}$ at a discrete set of time points $t_1, \dots, t_N$ to obtain the model output $g(\mathbf{x}(t_n), \hat{\mathbf{p}})$.
    \item Simulate $M$ synthetic datasets $\mathbf{y}_m$ by adding noise to the model output at each time point.
    \item For each synthetic dataset $\mathbf{y}_m$, estimate the best-fit parameter set $\mathbf{p}_m$ by minimizing the discrepancy between the model and the data.
    \item Compute the spread of the estimated parameters $\{\mathbf{p}_1, \dots, \mathbf{p}_M\}$.
\end{enumerate}

Although no universally accepted criterion exists \cite{lam2022practical}, a parameter is generally considered practically identifiable when its estimates remain sufficiently concentrated around the true value across the 
$M$ simulations. Spread can be quantified by metrics such as confidence-region volume, standard deviation, or average relative error \cite{lam2022practical, miao2011identifiability, tuncer2018structural}.

\subsection{Independent Gaussian Noise Simulation Method}

In the standard implementation of the MC approach, synthetic datasets are generated by solving the model deterministically and then adding Gaussian noise to the output. Specifically, for each time point $t_n$ and simulation $m$, the noisy observation is given by:
\begin{equation} \label{eq: data_sim}
    y_{n,m} = g(\mathbf{x}(t_n), \hat{\mathbf{p}}) \cdot (1 + \epsilon_{n,m}), \quad
    n = 1, \dots, N,\quad m = 1, \dots, M.
\end{equation}
where $\epsilon_{n,m} \sim \mathcal{N}(0, \sigma^2)$.
The scaled noise structure is a common choice for epidemic models, where uncertainty tends to grow with counts magnitude. An unscaled alternative may be more appropriate when measurement error is independent of case numbers, such as in repeatable laboratory assays \cite{miao2011identifiability}. 

The standard deviation $\sigma$ of the normal distribution from which the noise is sampled is commonly used as a threshold when evaluating identifiability. For example, we can compute the average relative estimation error:

\begin{equation}
    \text{ARE}(p^{(k)}) = \frac{1}{M} \sum_{m=1}^M \frac{|\hat{p}^{(k)} - p^{(k)}_m|}{\hat{p}^{(k)}} \times 100\%,
\end{equation}

\noindent where $p^{(k)}$ is the $k$-th parameter in the set $\mathbf{p}$ and compare it to the noise level $\sigma$ used to generate the synthetic data. Following the criteria in \cite{tuncer2018structural}, a parameter $p^{(k)}$ is considered \emph{practically identifiable} if its ARE remains below the noise level used to generate synthetic data. Common choices for this value include $\sigma = 0.05$, $0.1$, or $0.2$, corresponding to $5\%$, $10\%$, and $20\%$ noise levels \cite{chis2011structural, miao2011identifiability}. 

Although straightforward, this method assumes symmetric, time-invariant, and independent noise --- assumptions that rarely hold for epidemic data and, as we show in subsequent sections, fail to properly incorporate stochastic variability.

\subsection{The SIR model}
We consider the simple Susceptible-Infectious-Recovered (SIR) model framework as the foundation of this study. This model is described by the following system of ordinary differential equations:
\begin{equation}\label{eq:SIR}
\begin{split}
    \frac{dS}{dt} &= -\beta S  I\\
\frac{dI}{dt} &= \beta S I - \alpha  I \\
\frac{dR}{dt} &= \alpha  I
\end{split}
\end{equation}
Here, $S(t), I(t),$ and $R(t)$ are the number of susceptible, infectious, and recovered individuals at time $t$, respectively. We assume a closed population with no births or deaths. Hence, the total population $N(t) = S(t)+I(t)+R(t)$ is assumed to remain constant throughout the duration of the outbreak. The dynamics of the epidemic are governed by the transmission rate $\beta$ and the recovery rate $\alpha$. These two parameters also determine the basic reproduction number, which for this model is defined as $R_0 = \beta N / \alpha$. Throughout this study, we assume that the observed quantity is the prevalence curve $I(t)$; that is, the number of currently infectious individuals over time.

\subsection{CTMC Formulation} 
To simulate the inherent stochasticity of epidemic transmission, we use a continuous-time Markov chain (CTMC) formulation implemented via the Gillespie Algorithm \cite{gillespie1977exact}. We specifically implement the CTMC formulation for the SIR model. As prescribed by \cref{eq:SIR}, the only two types of events that can occur within this model are new infections and recoveries. We denote these events by $E_1$ and $E_2$, respectively. In the CTMC setting, time is treated as a continuous variable, and we assume that only one event occurs in any infinitesimally small time interval. 

The simulation begins by assigning initial conditions to each entry of the state variable vector $x = [S, I, R]$ and computing the transition rates $a_j(x)$ for each possible event $j$. In the case of the SIR model, $a_1(x) = \beta S I$ and $a_2(x) = \alpha I$. At each iteration $i$, the waiting time until the next event, denoted by $\tau_i$, is sampled from an exponential distribution with rate parameter $\sum_j a_j(x_i)$. The specific event $E_j$ that occurs during the interval $[t_i, t_i + \tau_i)$ is selected by sampling a uniform random variable \( u \sim \mathcal{U}(0,1) \) and comparing it to the cumulative sum of the transition probabilities. In the case of the SIR model, we select $E_1$ if $u < \frac{a_1(x_i)}{a_1(x_i) + a_2(x_i)}$, and $E_2$ otherwise.

After determining which event $E_j$ occurs, the time and state variables are updated so that $t_{i+1} = t_i + \tau_i$ and $x_{i+1} = x_i + m_j$, where $m_j$ is the $j^{\text{th}}$ row of the transition matrix $M$. This matrix encodes the change in state associated with each event, and for the SIR model it is defined as:

\[
M =
\begin{bmatrix}
-1 & 1 & 0 \\
0 & -1 & 1 \\
\end{bmatrix}.
\]

Although the CTMC simulation evolves in continuous time, for the purpose of analysis and comparison with ODE-based methods, we extract the state variables at discrete time points, specifically daily intervals. This yields time-series data suitable for parameter fitting and coverage analysis.

While computationally demanding, especially at high population counts, the CTMC approach serves as a high-fidelity reference for characterizing intrinsic stochastic variation in epidemic dynamics.

\subsection{Parameter Selection} \label{sec: par selection}
\begin{figure}[tbp]
    \centering
    \includegraphics[width=0.8\linewidth]{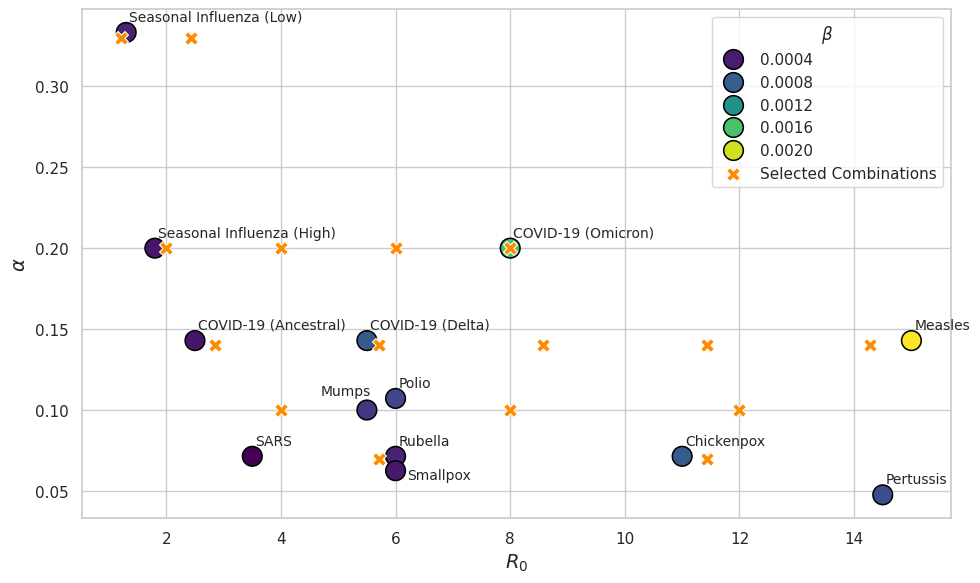}
    \caption{Mean recovery rate ($\alpha$) and basic reproduction number ($R_0$) values for a selection of infectious diseases. The transmission rates ($\beta$) are indicated by color shading and exact values are based on a total population of $N =1000$. Each point represents the average $(\alpha, R_0)$ values reported in the literature for each disease. Orange X markers indicate the 16 parameter combinations selected for data simulation and coverage analysis. We note that although some diseases included here may be better modeled with an SEIR framework, the $R_0$ values are the same for both SIR and SEIR formulations in the absence of natural or disease-induced mortality}
    \label{par_selection}
\end{figure}

In the analysis that follows, we explore a variety of realistic epidemic scenarios, as parameter identifiability and estimation uncertainty are known to vary significantly under different transmission dynamics, basic reproduction numbers, infectious durations, and growth rates \cite{eisenberg2013identifiability, pauer2009comparison, rohani2011importance}. To this end, we define a grid of plausible parameter combinations guided by values reported in the literature for real-world diseases \cite{arsal2020short, biggerstaff2014estimates, campbell2015pertussis, dye2003modeling, li2020early, liu2021reproductive, thompson2020review}. Specifically, we consider recovery rates $\alpha \in \{0.07, 0.1, 0.14, 0.2, 0.33\}$, corresponding to average recovery periods between 3 and 14 days. For each $\alpha$, we choose transmission rates $\beta$ such that the resulting basic reproduction number $R_0 = \beta N / \alpha$ spans approximately 1 to 15. For example, for a population of $N=1000$, we let $\beta = \{ 0.0004, 0.0008, 0.0012, 0.0016, 0.002\}$. From the full grid of 25 possible $(\alpha, \beta)$ combinations, we select 16 representative pairs that evenly span the $(\alpha, R_0)$ space and reflect the range of known dynamics for diseases such as seasonal influenza, measles, COVID-19, and SARS. These selected parameters are plotted along with known disease parameters in \cref{par_selection}.

\section{Characterizing Stochastic Noise}\label{sec:stochastic_analysis}
In this section, we explore the structure of stochastic noise for a variety of parameter combinations to understand whether it can be embedded within independent Gaussian noise, quantify the extent and importance of time dependence, and identify systematic patterns in relation to $R_0$.
We begin by analyzing the statistical properties of residuals between CTMC trajectories and the deterministic ODE solution, with a focus on asymmetry, dispersion, and temporal dependence. We then evaluate whether non-Gaussian distributions with time-varying dispersion can provide a more realistic approximation of the underlying stochastic variation. Finally, we implement an empirical sampling approach that resamples residuals directly from CTMC trajectories to evaluate whether, in principle, a more faithful noise distribution can capture stochastic behavior, without temporal dependence. We assess the degree to which each noise model captures stochastic variability by calculating its coverage probability, as presented in \cref{sec:coverage_results}

\subsection{Residual Structure Analysis}
\begin{figure}[tbph]
    \centering
    \includegraphics[width=0.95\linewidth]{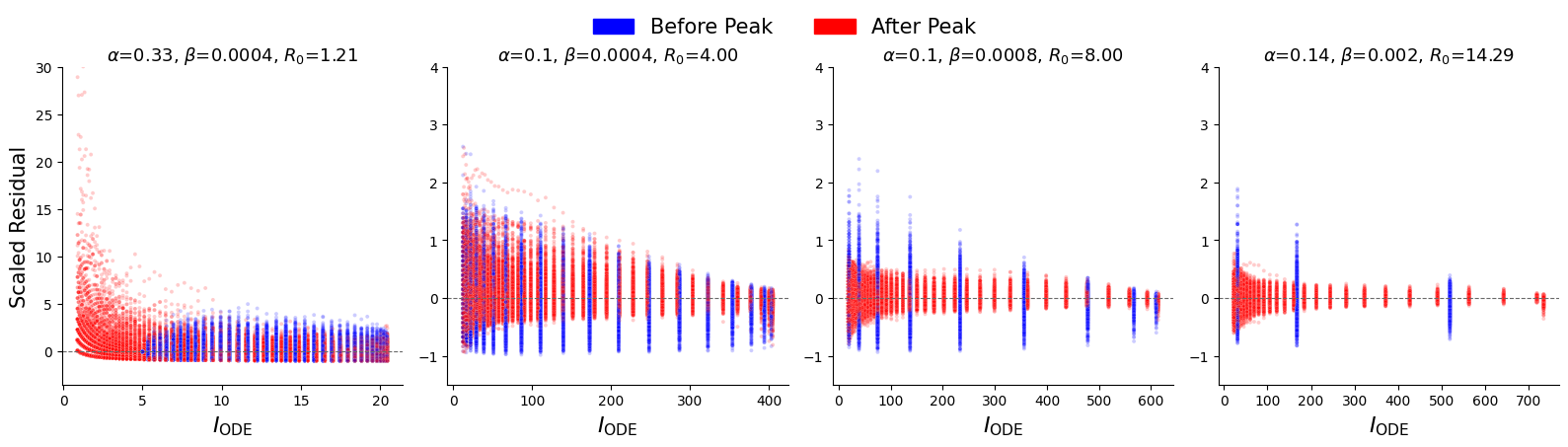}
    \caption{Scatter plots of scaled residuals vs. model-predicted prevalence for representative $(\alpha, \beta)$ parameter combinations, sorted by increasing basic reproduction number. Each point represents the residual from a CTMC trajectory at a given time, with color indicating whether the point occurs before (blue) or after (red) the epidemic peak. The left most panel uses an extended vertical axis limit to accommodate extreme outliers observed in low $R_0$ settings.}
    \label{fig:residual-structure}
\end{figure}

To analyze the structure of stochastic deviations, we simulate $J=1000$ CTMC trajectories for each parameter combination selected in \cref{sec: par selection} and analyze the distribution of scaled residuals $\epsilon_n(t) = \frac{\text{CTMC}_n(t) - I(t)}{I(t)}$,
where $I(t)$ is the solution of the deterministic ODE system. 
\Cref{fig:residual-structure} provides a visual overview of residual behavior across a few representative parameter combinations. A complete plot of all 16 combinations as well as residuals vs time plots are shown in \cref{appendix:additional_figs}. At low $R_0$ values, residuals are highly skewed and dispersed, especially during early epidemic growth. As $R_0$ increases, residuals become more symmetric and concentrated around zero. Moreover, there is a clear dependence on epidemic phase, with residuals generally more spread out before the epidemic peak and more densely packed afterward. This pattern is consistent with the greater degree of uncertainty that is usually observed in the early stages of an outbreak, when case counts are low and stochastic fluctuations are high. Overall, these trends suggest that stochastic noise is both phase- and magnitude-dependent, asymmetric, and time-varying.

\begin{figure}[tbph]
    \centering
    \includegraphics[width=0.5\linewidth]{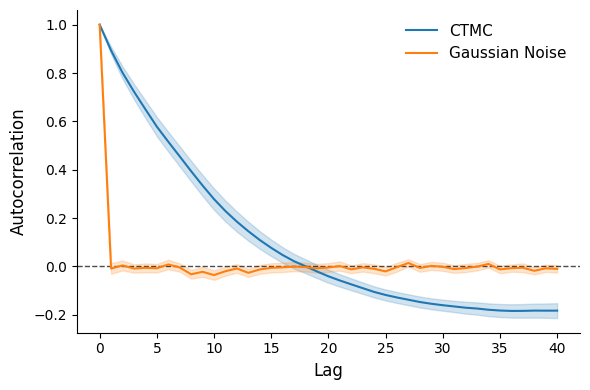}
    \caption{Autocorrelation plots for residuals from CTMC simulations (blue) and from independent Gaussian noise added to the ODE solution (orange). Each curve represents the mean autocorrelation function (ACF) across 100 trajectories for a fixed parameter set ($\alpha = 0.1$, $\beta = 0.0004$), with shaded bands indicating the 95\% confidence interval across simulations.}
    \label{fig:autocorrelation}
\end{figure}

To further investigate temporal structure, we generate autocorrelation plots for residuals from CTMC trajectories and from independent Gaussian noise added to the ODE output (see \cref{fig:autocorrelation}). The slow decay towards zero of the autocorrelation function for the CTMC residuals indicates strong time dependence. In contrast, Gaussian noise residuals show no temporal correlation, as expected under an independence assumption. While subsampling the time series could reduce correlation, it would also lead to information loss and, as demonstrated in prior work \cite{saucedo2024comparative}, significantly reduce parameter identifiability. Additionally, a variance vs. mean count plot (see \cref{fig:poiss} in \cref{appendix:additional_figs}) shows that variance often exceeds the mean, especially prior to the epidemic peak, indicating super-Poisson behavior. These results suggest that stochastic variation in epidemic dynamics is not only asymmetric and phase-dependent, but also temporally structured and overdispersed. While this highlights the complexity of stochastic noise, it is not immediately clear which features are essential for reproducing its effects. In the next section, we test whether capturing marginal characteristics such as skewness and overdispersion—while still ignoring temporal dependence—is sufficient to approximate the observed variability.


\subsection{Fitting Parametric Noise Models}
To evaluate whether temporal dependence can be ignored while still preserving realistic stochasticity, we fit several candidate distributions to the residuals. Specifically, we test the normal, log-normal, gamma, Weibull, and skew-normal distributions. These are selected to cover a variety of properties relevant to epidemic modeling, such as skewness and varying degrees of dispersion.  Since some of these distributions have strictly positive support, we shift the residuals by +1 to ensure non-negativity before fitting (raw residuals are naturally bounded below by $-1$, corresponding to a prevalence count of zero).

\begin{figure}[tbph]
    \centering
 \includegraphics[width=0.9\linewidth]{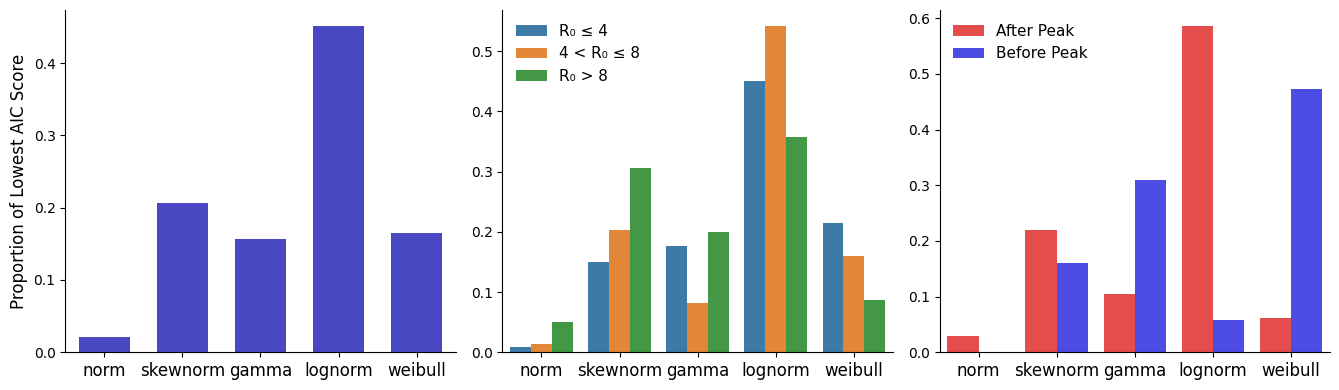}
    \caption{Comparison of the best-fitting distributions for scaled residuals based on Akaike Information Criterion (AIC). On the left panel, we  show the overall frequency with which each candidate distribution achieves the lowest AIC score across all parameter combinations and prevalence counts. On the middle and right panels we plot the proportion of AIC wins stratified by basic reproduction number and epidemic phase, respectively.}
    \label{fig:AIC}
\end{figure}

We compare the fit of each candidate distribution using the Akaike Information Criterion (AIC), with parameters estimated via maximum likelihood using \texttt{scipy.stats.fit}. The left panel of \cref{fig:AIC} displays the frequency with which each distribution achieves the lowest AIC score across prevalence counts and $(\alpha, \beta)$ combinations. The log-normal distribution is the most frequently selected model, followed by the skew-normal and Weibull distributions. Additionally, we stratify AIC comparisons by basic reproduction number and epidemic phase in order to assess whether these trends manifest across different transmission scenarios. As shown in the middle panel of \cref{fig:AIC}, the log-normal distribution remains the best-fitting model across all $R_0$ levels, especially for low and moderate transmission scenarios.
When stratifying by epidemic phase, the log-normal distribution achieves the highest proportion of best-fit selections \textit{after} the epidemic peak, but is outperformed by other distributions in the \textit{pre-peak} phase, where the Weibull distribution performs best (right panel of \cref{fig:AIC}). Notably, the normal distribution is selected the least frequently in all groups, reflecting its inability to capture the skewness and heavy tails present in the residuals.

To evaluate goodness-of-fit of the log-normal distribution beyond relative AIC rankings, we employ the Anderson--Darling (A--D) test. This test is flexible in cases where distributional parameters are estimated from the data itself, as in our setting. We first apply a log-transformation to the residuals and then apply the A--D test with the corrected test statistic for the normal distribution provided in \cite{stephens1974edf}. 

\begin{figure}[tbph]
     \centering
\includegraphics[width=0.5\linewidth]{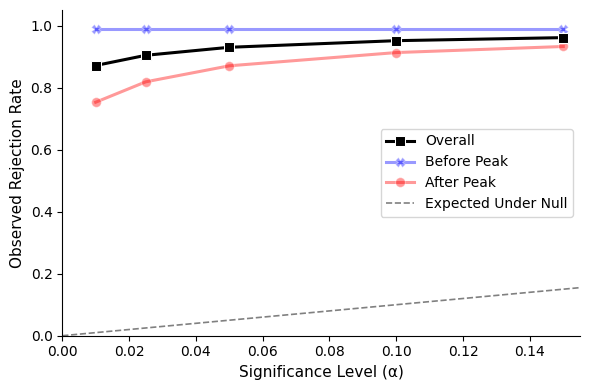}
     \caption{Observed rejection rates from the Anderson--Darling test applied to log-transformed residuals. Rejections for residuals before the peak are shown in blue, after the peak in red, and for both phases combined in black. The dashed diagonal line, $y = \alpha$, represents the expected rejection rate under the null hypothesis. }
     \label{fig:ad-rejection}
\end{figure}

As illustrated in \cref{fig:ad-rejection}, the null hypothesis of log-normality is rejected at a high rate across all significance levels tested. For example, rejection rates exceed 90\% at the conventional 5\% significance level. Residuals from  \textit{before-peak} have rejection rates close to 1 even at higher thresholds, indicating a particularly poor fit in the early stages of the epidemic, as was already suggested by the AIC test. Residuals from \textit{after-peak} show a slightly better fit to the log-normal model, although rejection rates remain high. These results suggest that, despite being the best distribution among the ones tested, the log-normal distribution fails to capture key characteristics of the residual structure.

\subsection{ Empirical Noise Simulation}\label{sec: empirical_noise}
\begin{figure}
    \centering
    \includegraphics[width=0.8\linewidth]{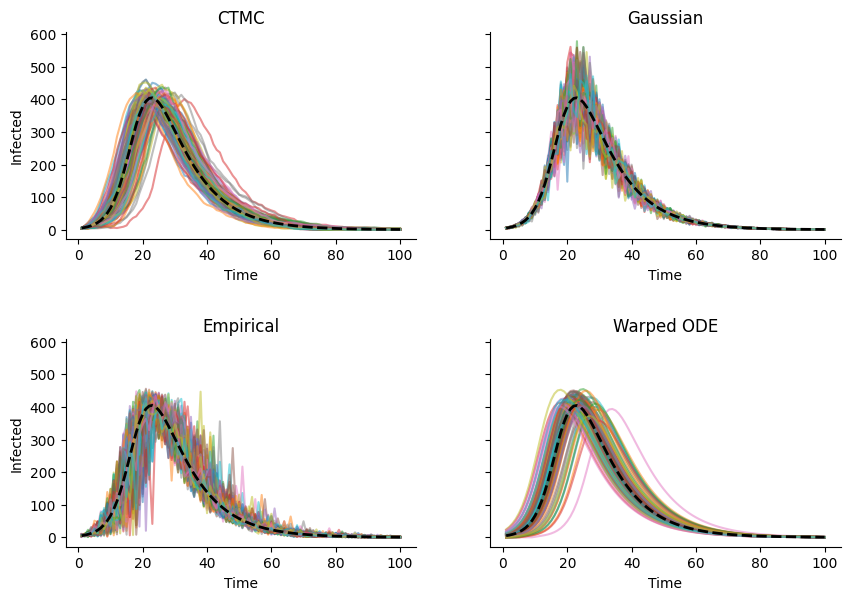}
    \caption{
    Comparison of simulated epidemic trajectories under four data-generating mechanisms: CTMC simulations, Gaussian noise ($\sigma = 0.1$) added independently to the ODE output, empirical residual resampling, and amplitude/time-warped ODE curves. The black dashed line represents the deterministic ODE solution. Simulations were performed with $N = 1000$, $\alpha = 0.1$, and $\beta = 0.0004$.
    }
    \label{fig:traj_comparison}
\end{figure}

Given that none of the distributions we tested provide a good fit to the residuals, we consider a fully nonparametric approach by resampling directly from the empirical residuals. Residuals are stratified by prevalence bin, epidemic phase, and parameter combination. Synthetic data sets are generated by independently sampling residuals from the appropriate bin and adding them to the ODE solution following the structure in \cref{eq: data_sim}. Although this approach is not feasible in practical applications, where the true residual distribution is unknown, it serves as a proof of concept as it allows us to evaluate whether adding noise \textit{independently} sampled from a well-approximated distribution is able to reproduce the stochastic variability underlying an epidemic spread. While visually this method reproduces the envelope of uncertainty around the ODE solution, as shown in \cref{fig:traj_comparison}, it fails to replicate the structured deviations seen in CTMC trajectories. As shown in \cref{tab:coverage_all}, coverage is significantly lower than the set confidence level across all parameter combinations, suggesting that the lack of temporal dependence in the synthetic noise remains a critical limitation. These results highlight that even well-fitted marginal noise models cannot reproduce key features of stochastic variation. In the following section, we introduce a new strategy that addresses this limitation by introducing time-dependent variability directly into the ODE solution.


\section{Time-Dependent Noise via Warping and Amplitude Scaling}\label{sec: hybrid_method}

Despite being empirically informed, the noise model in \cref{sec: empirical_noise} remains limited by its core assumption of temporal independence. This assumption fails to capture the temporal correlations and structural variability observed in CTMC simulations, such as variation in peak timing and height, or duration of the epidemic. To address this, we develop a hybrid simulation approach aimed at approximating stochastic variability more realistically than independent noise models, while avoiding the computational burden of simulating full CTMC trajectories. Instead of simulating a full stochastic process, we transform the deterministic ODE solution to replicate key sources of variability observed in CTMC models.

Our approach draws inspiration from the literature on functional data alignment, particularly \textit{curve registration} \cite{ferraty2011oxford, ramsay1998curve}, where the goal is to align salient curve features of a collection of functional data to a common template. An illustrative example is shown in \cref{appendix:curve_registration}, where we apply curve registration to CTMC trajectories and find that the registered mean closely matches the ODE solution. Here, we reverse the logic: starting from a deterministic ODE solution, we generate new trajectories by applying transformations that mimic the types of variability observed in stochastic simulations. As discussed in detail in \cite{ferraty2011oxford}, functional variation can often be decomposed into two key components: \textit{amplitude} variation, which alters the vertical scale of the trajectory, and \textit{phase} variation, which shifts its timing. Hence, we define a synthetic infectious trajectory as:
\begin{equation}
    \tilde{I}(t) = a \cdot I_{\text{ODE}}(t + \Delta t)
\end{equation}
where \( a \) is an amplitude scaling factor and \( \Delta t \) is a time shift applied to the ODE solution \( I_{\text{ODE}}(t) \).  Since the shifted time points \( t + \Delta t \) may not align with the original evaluation grid, we compute \( I_{\text{ODE}}(t + \Delta t) \) using linear interpolation via \texttt{scipy.interpolate.interp1d}. This allows us to evaluate the ODE solution at arbitrary time shifts and ensures continuity in the resulting warped trajectories.

To test the validity of this simulation strategy, we fit a kernel density estimator to the joint empirical distribution of $a$ and $\Delta t$ values extracted from CTMC simulations, and sample new shift–scale pairs from this smoothed distribution. Given the observed negative correlation between peak intensity and time, we sample the two parameters jointly. As we demonstrate in Section~\ref{sec:coverage_results}, this method achieves markedly improved coverage compared to temporally independent noise models, and its performance is comparable to that of full CTMC simulations.
Visually, we can see that applying this transformation to the ODE solution generates time series that more closely resemble the range of CTMC trajectories in both shape and spread, as shown in \cref{fig:traj_comparison}. This method retains the computational efficiency of ODE-based simulations while offering a flexible way to introduce temporally structured noise. 

It is well established that the extent of stochastic variation in epidemic dynamics depends on both the basic reproduction number $R_0$
and the population size $N$ \cite{allen2008introductio}. Lower $R_0$ values are associated with smaller outbreak sizes and a greater influence of stochastic fluctuations, while larger populations tend to smooth out individual-level randomness due to the law of large numbers. Accordingly, the magnitude of amplitude and time shifts should be informed by the underlying transmission setting. To quantify the amount of stochastic variation present across scenarios and guide the choice of $\Delta t$ and $a$, we calculate the empirical mean and standard deviation of both quantities from CTMC simulations. These statistics are computed across the 16 parameter combinations and three population sizes, $N = 100, 1000$ and $10,000$. These values were chosen to reflect three representative scenarios, in terms of the effective population size to which individuals are exposed. A population of 100 resembles a small, confined environment such as the 1978 influenza outbreak in a British boarding school \cite{center1978influenza}; 1000 represents a medium-scale population such as a small town or tightly connected community; and 10,000 approximates a large-scale urban subpopulation within a city, such as a neighborhood or school district.

\begin{figure}[tbph]
    \centering
    \includegraphics[width=0.8\linewidth]{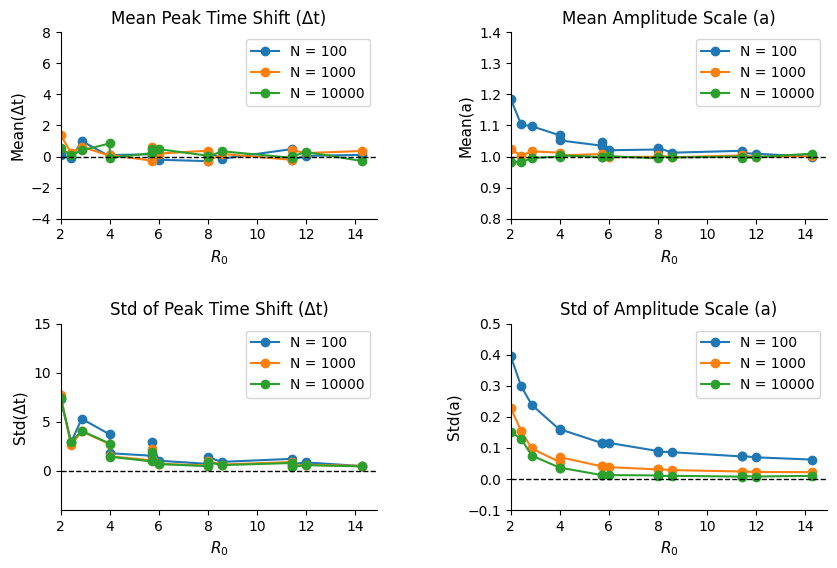}
    \caption{Dependence of peak time shift $\Delta t$ and amplitude scaling factor $a$ on the basic reproduction number $R_0$ and  population size $N$.}
    \label{fig:shifts}
\end{figure}

As shown in \cref{fig:shifts}, the mean time shift stays approximately around zero across $R_0$ and $N$ values. Its standard deviation is mostly independent of $N$ but is higher at lower $R_0$ and declines toward zero as $R_0$ increases. The amplitude scaling factor has mean around one across almost all simulation scenarios except at low $R_0$ and small $N$, where upward bias is observed. 
Its standard deviation shows clear dependence on both $R_0$ and $N$, with values close to zero for large population counts and $R_0$ above six and higher values at lower $R_0$ and lower population counts. Together, these findings confirm that the strength of stochastic effects is highly scenario-dependent and provide a principled basis for selecting the amount of amplitude and timing variability to be applied into ODE-based simulations. These results also suggest that, in high-transmission, large-population settings, where stochastic variation is minimal, independent Gaussian noise may remain an adequate approximation. However, modeling stochasticity more faithfully becomes essential in more variable scenarios, such as small populations or low $R_0$.

\section{Coverage Assessment}\label{sec: coverage}
\subsection{Coverage Computation Protocol}
In order to evaluate how well each  simulation method described in \cref{sec:stochastic_analysis} captures stochastic variability, we compute its \textit{coverage probability}. This quantifies how often a confidence region constructed from simulated data contains the true parameter values. To compute coverage we execute the following steps:

\begin{enumerate}
    \item Generate $J = 100$ stochastic realizations from the true parameter set $\hat p = (\hat \alpha, \hat \beta)$.
    \item Fit the ODE model to each realization and compute the best-fit parameter pair $(\alpha_j, \beta_j)$ for $j = 1, \ldots, J$.
    \item For each fitted pair $(\alpha_j, \beta_j)$, generate $M = 1000$ synthetic datasets $y_j^m$ using a chosen simulation method.
    \item For each simulated dataset $y_j^m$, fit the model and recover the estimated parameters $(\alpha_j^m, \beta_j^m)$.
    \item For each $j$, construct a joint $1 - \alpha_c$ confidence region using the $M$ parameter estimates.
    \item Evaluate whether the true parameter vector $\hat p$ falls within the estimated confidence region for each of the $J$ trials.
\end{enumerate}
Coverage is then computed as the proportion of times (out of $J$) that the true parameters fall within their respective confidence regions. A method is said to have good coverage if this proportion closely matches the target confidence level $1- \alpha_c$.

\subsection{Coverage Results}\label{sec:coverage_results}

We evaluate the coverage of each simulation method described earlier for each of the 16 parameter combinations selected in \cref{sec: par selection}. For each method, we compute the coverage by following the steps delineated above. To construct joint confidence regions for each method, we apply a kernel density estimate (KDE) to the 2D distribution of estimated parameters. These KDEs define the $1 - \alpha_c$ confidence contour used to calculate coverage which we show in \cref{tab:coverage_all}. In this example, we set $\alpha_c = 0.32$ to target a nominal $68\%$ level, which serves as a representative threshold for evaluating relative differences across methods. As expected, the CTMC-based method achieves consistently good coverage across all parameter sets. The naive Gaussian-noise simulation approach, where independent noise is added at each time point, performs poorly across nearly all parameter combinations. This underperformance is especially pronounced when $R_0$ is low, where transmission is slower and stochastic fluctuations play a larger role. The empirical noise method also fails to achieve good coverage, indicating that its approximations still miss critical features of the underlying noise. In contrast, the hybrid warped model performs much better, attaining coverage performance comparable to that of the CTMC at a significantly reduced computational cost. 

\begin{table}[htbp]
    \centering
    \begin{tabular}{ccc|ccccccc}
        \toprule
        $\alpha$ & $\beta$ & $R_0$ & CTMC &
        \makecell{Gaussian\\($\sigma = 0.1$)} &
        \makecell{Gaussian\\($\sigma = 0.2$)} &
        \makecell{Empirical\\Noise} &
        \makecell{Hybrid} \\
        \midrule
        0.33 & 0.0004 & 1.21 & 0.77 & 0.00 & 0.01 & 0.02  & 0.79\\
        0.20 & 0.0004 & 2.00 & \cellcolor{yellow!30}0.75 & 0.01 & 0.04 & 0.05 & 0.78\\
        0.33 & 0.0008 & 2.42 & \cellcolor{yellow!30}0.67 & 0.08 & 0.12 & 0.01 & \cellcolor{yellow!30}0.70\\
        0.14 & 0.0004 & 2.86 & \cellcolor{yellow!30}0.74 & 0.05 & 0.17 & 0.02  & \cellcolor{yellow!30}0.72\\
        0.10 & 0.0004 & 4.00 & \cellcolor{yellow!30}0.67 & 0.04 & 0.19 & 0.04  & \cellcolor{yellow!30}0.75\\
        0.20 & 0.0008 & 4.00 & \cellcolor{yellow!30}0.67 & 0.06 & 0.28 & 0.02   & \cellcolor{yellow!30}0.67\\
        0.07 & 0.0004 & 5.71 & \cellcolor{yellow!30}0.63 & 0.12 & 0.30 & 0.03  & \cellcolor{yellow!30}0.73\\
        0.14 & 0.0008 & 5.71 & \cellcolor{yellow!30}0.75 & 0.18 & 0.36 & 0.04  & 0.79\\
        0.20 & 0.0012 & 6.00 & \cellcolor{yellow!30}0.64 & 0.17  & 0.45 & 0.01   & 0.60\\
        0.10 & 0.0008 & 8.00 & \cellcolor{yellow!30}0.63 & 0.16 & 0.42 & 0.00  & \cellcolor{yellow!30}0.71\\
        0.20 & 0.0016 & 8.00 & \cellcolor{yellow!30}0.63 & 0.24 & 0.55 & 0.04  & \cellcolor{yellow!30}0.63\\
        0.14 & 0.0012 & 8.57 & \cellcolor{yellow!30}0.64 & 0.19 & 0.47 & 0.02 & \cellcolor{yellow!30}0.65\\
        0.07 & 0.0008 & 11.43 & \cellcolor{yellow!30}0.69 & 0.24 & 0.51 & 0.03 & \cellcolor{yellow!30}0.69 \\
        0.14 & 0.0016 & 11.43 & \cellcolor{yellow!30}0.66 & 0.28 & 0.58 & 0.00 & \cellcolor{yellow!30}0.72\\
        0.10 & 0.0012 & 12.00 & \cellcolor{yellow!30}0.63 & 0.19 & 0.49 & 0.05 & 0.\cellcolor{yellow!30}61\\
        0.14 & 0.0020 & 14.29 & \cellcolor{yellow!30}0.64 & 0.26 & \cellcolor{yellow!30}0.74 & 0.01 & \cellcolor{yellow!30}0.66\\
        \bottomrule
    \end{tabular}
    \caption{Comparison of coverage probabilities across simulation methods and parameter settings. Coverage reflects the proportion of times the 68\% joint confidence region contained the true parameter vector. Cells highlighted in yellow indicate coverage within $\pm 10 \%$ of the nominal confidence level.}
    \label{tab:coverage_all}
\end{table}

\begin{figure}[tbph]
    \centering
    \includegraphics[width=0.9\linewidth]{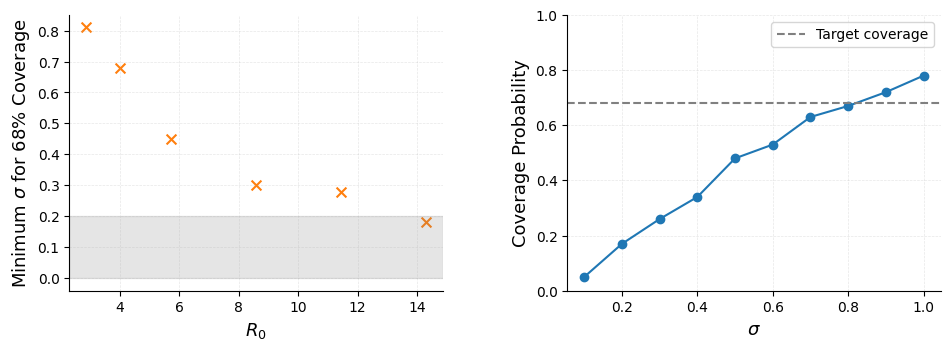}
    \caption{Coverage analysis for the naive noise model. On the left panel we plot the estimated minimum noise level ($\sigma_{min}$) required to reach the nominal $68\%$ coverage across a subset of parameter combinations, plotted against $R_0$. The shaded gray region highlights typical ranges for realistic measurement noise. On the right, we plot a full $\sigma$-coverage curve for a representative parameter set ($\alpha = 0.14, \beta = 0.0004$).}
    \label{fig:coverage}
\end{figure}

To better understand the limitations of the naive method, we conducted additional analyses to explore how increasing the noise level affects coverage. First, we estimated the minimum noise level ($\sigma_{min}$) required for the naive simulation approach to reach $\approx 68\%$ coverage. We do this for a subset of six representative parameter combinations that span a variety of $R_0$ values. As shown in the left panel of \cref{fig:coverage}, this required $\sigma_{min}$ decreases with increasing $R_0$, but remains unrealistically high across the board. This indicates that large unstructured noise levels are required to approximate the uncertainty seen in stochastic simulations --- an approach that distorts the dynamics rather than replicating them. Additionally, on the right panel of \cref{fig:coverage}, we provide an example of a full coverage curve across a range of increasing values of $\sigma$. We observe that coverage increases linearly with noise, but only reaches acceptable levels when $\sigma$ exceeds $\approx 0.8$, which once again far exceeds realistic assumptions for measurement noise.

\section{Identifiability Comparison}\label{sec:identifiability}
\begin{figure}
    \centering
    \includegraphics[width=0.9\linewidth]{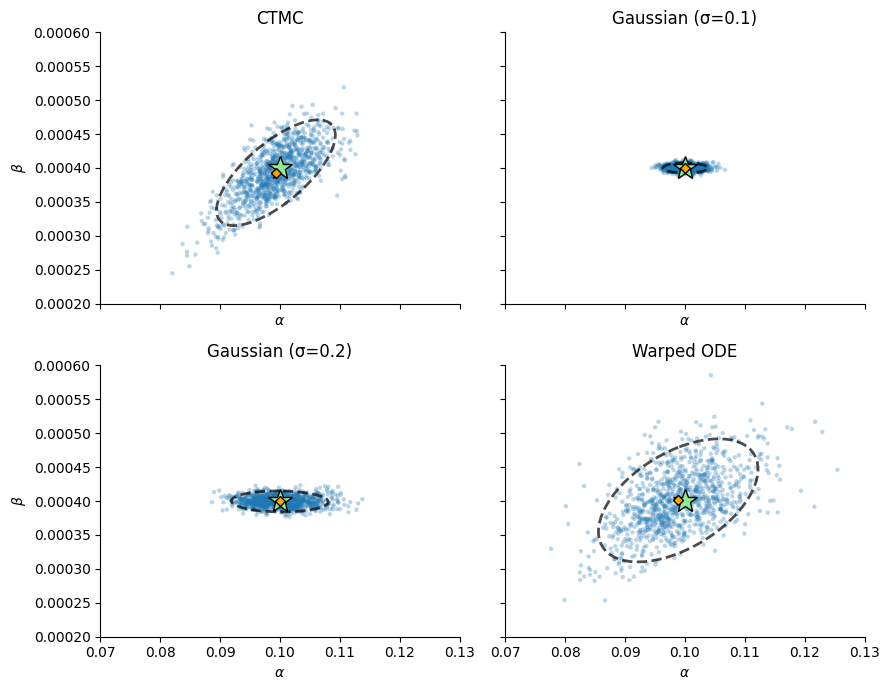}
    \caption{Scatter plots of best-fit parameter estimates \(\hat{\alpha}\) and \(\hat{\beta}\) obtained from fitting 1000 simulated trajectories using different noise models: CTMC (top left), Gaussian noise with \(\sigma = 0.1\) (top right), Gaussian noise with \(\sigma = 0.2\) (bottom left), and warped ODE simulations (bottom right). The green star and orange dot mark the true and mean parameter values, respectively. Dashed ellipses denote the $95\%$ empirical confidence regions of the parameter estimates, assuming approximate joint normality.}
    \label{fig:spread}
\end{figure}

To complement the coverage analysis, we evaluate how simulation method impacts practical identifiability by comparing the spread of best-fit parameter estimates under each method. This allows us to assess not only how well each method reproduces uncertainty, but also how it may affect inference about parameter identifiability. For every method, we generate $J = 1000$ synthetic datasets from the true parameter vector $\hat{p} = (\hat\alpha, \hat\beta)$ and compute the best-fit parameters by minimizing the squared error between the simulated trajectory and the deterministic model solution. 
\Cref{fig:spread} shows scatter plots of the resulting estimates, and a quantitative summary of their dispersion is provided in \cref{tab:spread}, which reports the coefficient of variation (CV) of each estimated parameter.
A key observation is that stochasticity itself, as introduced by the CTMC simulation method, leads to significantly greater dispersion in the recovered parameters compared to even relatively high levels of Gaussian noise. The substantial differences in CV values between the CTMC and Gaussian models, particularly for $\beta$, have important implications for practical identifiability analysis, which often relies on the spread of best-fit estimates as an assessment metric. For example, the markedly lower CVs under Gaussian noise suggest that process noise alone can produce considerable identifiability challenges, and that additive Gaussian noise may underrepresent the true parameter variability introduced by uncertainty observed in real outbreaks. Notably, the warped ODE approach produces a spread in parameter estimates that closely resembles the CTMC in shape but tends to be slightly wider, which may reflect a modest overestimation of uncertainty under this method. Nevertheless, its overall similarity further reinforces its ability to approximate realistic stochastic variation.


\begin{table}[tbph]
\renewcommand{\arraystretch}{1.2}
\centering
\begin{tabular}{l | c | c }
\hline
   & Coefficient of Variation of $\alpha$ &  Coefficient of Variation of $\beta$ \\ \hline 
 CTMC  & $4.77\%$ &  $9.64\%$\\ 
 Gaussian ($\sigma = 0.1$) & $1.90\%$ &  $1.00\%$ \\  
  Gaussian ($\sigma = 0.2$) & $3.86\%$& $2.01\%$ \\  
  Hybrid & $6.46\%$& $11.19\%$\\  
   \hline
\end{tabular}
\caption{Coefficient of variation (CV) of best-fit parameter estimates across simulation methods. The CV is computed as the standard deviation divided by the mean, multiplied by 100.}
\label{tab:spread}
\end{table}


\section{Discussion}\label{sec:discussion}
Assessing the practical identifiability of model parameters is an integral part of the parameter estimation process, especially when the parameter estimates are used to make predictions, evaluate control strategies, or provide epidemiological interpretation. Monte Carlo methods offer an intuitive simulation-based approach for evaluating practical identifiability, relying on synthetic datasets to assess the uniqueness of parameter estimates under real-life variability. However, standard implementations of the MC method were first developed to use in experimental or lab settings, where measurements can often be repeated under controlled conditions and measurement error is the dominant source of uncertainty \cite{iso1993guide}. In such settings, adding independent Gaussian noise to a deterministic model trajectory can be an appropriate way to simulate synthetic data. 

In recent years, this framework has been applied to epidemic modeling, where repeatable observations are not available and uncertainty arises not only from measurement noise but also from intrinsic stochasticity in the transmission process. As highlighted in recent reviews \cite{lam2022practical, wieland2021structural}, the fidelity of the artificially simulated noise to real-world uncertainty has been brought into question and has been mentioned as an important limitation of the standard MC algorithm. In practice, the epidemic curve we observe is often a single realization of a stochastic process, shaped by randomness in infection and recovery events \cite{alahakoon2022estimation, britton2009epidemic}. Accurately capturing this variability is essential for reliable assessment of practical identifiability.

In this study, we systematically evaluated the extent and structure of stochastic variation in the SIR model across a range of plausible epidemiological scenarios. Our goal was to understand whether the assumptions underlying standard MC simulations --- in particular, the use of independent Gaussian noise --- are sufficient to model the types of uncertainty seen in epidemic dynamics, and how these choices affect identifiability conclusions.

We began by analyzing the structure of CTMC residuals and found that there is a strong dependence on time, epidemic phase, and $R_0$, with residuals exhibiting skewness and time-varying spread, especially in the early stages of an outbreak and at lower $R_0$ values. We then tested whether these features could be captured by alternative models that relax the assumptions of Gaussian noise with constant variance. While flexible parametric distributions like the log-normal provided a better fit than the standard Gaussian model, overall goodness-of-fit was poor across parameter combinations. Moreover, we demonstrated that even adapting an empirical resampling strategy still led to poor coverage in the absence of temporal structure, which highlights that the lack of time dependence remained a critical limitation. 

To address this, we proposed a hybrid simulation strategy that introduces time-dependent variability by applying stochastic time and amplitude shifts to the deterministic ODE trajectory. With this method, we aimed to preserve the computational efficiency and simplicity of ODE-based approaches while capturing the structured deviations observed in CTMC simulations. Despite being a simplification of the full stochastic process, this approach achieved good coverage across all transmission scenarios tested and visually reproduced the shape and spread of stochastic trajectories more faithfully than independent noise models. Although our implementation samples shift parameters from CTMC-derived distributions, we also analyzed how these shifts vary with transmission dynamics and population size, providing general guidelines for adapting the method to new settings.

We then turned to the implications for identifiability analysis. By comparing the spread of best-fit parameters across simulation methods, we found that stochastic simulations produced substantially greater dispersion than Gaussian-based approaches, even at relatively high noise levels. The warped ODE method yielded a similar spread magnitude and shape to the CTMC. Instead, independent Gaussian models consistently underestimated parameter variability, which can lead to overly optimistic conclusions about identifiability.
This also raises the risk that a model–data combination may be mistakenly deemed practically identifiable when using an independent Gaussian noise model, especially if coverage is not taken into consideration.

Overall, this study highlights that the standard implementation of the Monte Carlo algorithm for assessing practical identifiability may not adequately represent the true uncertainty present in epidemic outbreaks. In fact, while independent Gaussian noise may offer a convenient approximation in large populations with fast transmission dynamics, it significantly underestimates uncertainty in more stochastic settings --- such as small populations or low $R_0$ outbreaks where intrinsic variation plays a dominant role. In these cases, the noise model should explicitly account for stochastic variation, either through CTMC simulations (when computational resources and expertise permit) or through the hybrid approach we propose. Although our current implementation relies on CTMC-informed shift distributions, future work could focus on generalizing this step, for example by training a neural network to predict time and amplitude shifts based on population size and transmission characteristics.

\section*{Acknowledgments}
This work was supported in part by National Science Foundation grant DMS-2045843.

\bibliographystyle{unsrt}
\bibliography{References}

\appendix

\crefalias{section}{appendix}

\section*{Supplementary Material}
\addcontentsline{toc}{section}{Supplementary Material}

\section{Effect of Practical Unidentifiability on Control Strategy Evaluation}
\label{appendix:control_strategy}
\begin{figure}[ht]
    \centering
    \includegraphics[width=0.9\linewidth]{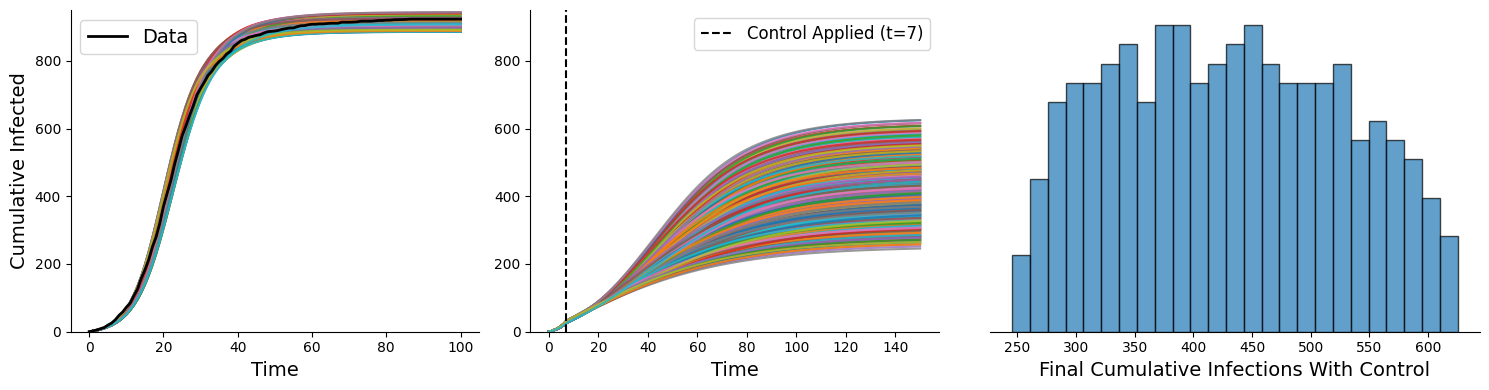}
    \caption{Left panel: plot of 100 best-fit ODE trajectories to the cumulative incidence data (black). Each trajectory corresponds to a different parameter pair $(\alpha_j, \beta_j)$ that provides a similarly good fit to the observed data. 
    Middle panel: the same 100 parameter sets are used to project the epidemic under a control intervention that reduces the transmission rate $\beta$ by 60\% at $t = 7$ days. The x-axis is extended to 150 days to accommodate the slower transmission under the control intervention.
    Right panel: histogram of the final cumulative incidence counts under the control strategy, showing substantial variability in projected epidemic size.}
    \label{fig:control}
\end{figure}

To illustrate the real-world consequences of practical unidentifiability, we present an  example where parameter uncertainty undermines the accurate evaluation of a proposed public health intervention. We consider the SIR model along with cumulative incidence data for a population of size $N=1000$. We aim to test the efficacy of a hypothetical control intervention that reduces the transmission rate $\beta$ by 60\% one week after the first observed case, simulating the effects of interventions such as social distancing or vaccination. 
Due to the unidentifiability inherent to this model-data combination, we are able to find a collection of parameter sets $(\alpha_j, \beta_j)$, $j = 1,..., 100$, each of which provides a similarly good fit to the observed data. In the left panel of \cref{fig:control}, we plot the ODE solutions corresponding to these parameter sets along with the cumulative incidence data.
Next, we apply the proposed intervention strategy and compare the predicted epidemic curves for each parameter set $(\alpha_j, \beta_j)$. As shown in the middle panel of \cref{fig:control}, despite applying the same control rule to each best-fit parameter set, the projected outcomes vary significantly from a slight reduction to one close to $75\%$ in the final count of cumulative incidence cases. A histogram of the final cumulative incidence counts under the control strategy is plotted in the right panel of \cref{fig:control}

This example highlights how unidentifiability of the model parameters can compromise downstream decision-making. Even when models fit data well, they may fail to provide reliable guidance for policy if parameter estimates are not uniquely determined.

\newpage
\section{Curve Registration} \label{appendix:curve_registration}

\begin{figure}[ht]
    \centering
    \includegraphics[width=0.9\linewidth]{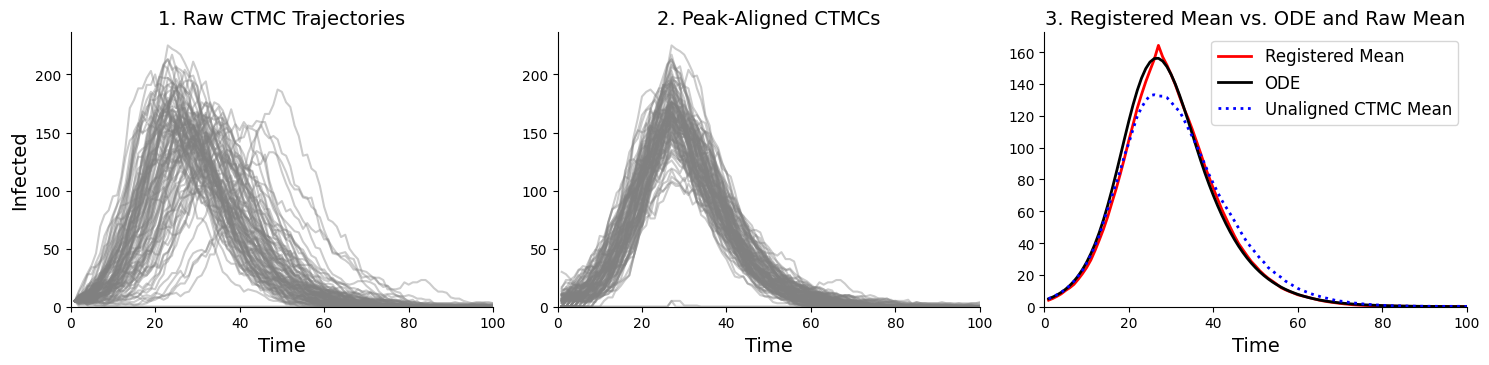}
    \caption{Forward curve registration using peak alignment for CTMC trajectories $(\alpha = 0.2, \beta = 0.0004, N=1000)$. The left panel shows raw CTMC trajectories, which exhibit substantial variability in both peak timing and height. The middle panel shows trajectories after peak alignment, which removes phase variation. The right panel compares the mean of aligned curves (red) with the ODE solution (black) and the unaligned mean (blue). Peak alignment recovers the ODE mean almost exactly, illustrating that much of the difference between CTMC and ODE means is due to phase variability rather than changes in overall shape.}
    \label{fig:curve_registration}
\end{figure}

The hybrid method we propose in this study is motivated by curve registration, a class of techniques in functional data analysis designed to align a collection of functions by removing phase variability. In this framework, curves that share similar overall shapes but differ in the timing of key features can be adjusted to a common reference. A common approach is \textit{landmark alignment}, in which curves are synchronized at identifiable features—for instance, CTMC prevalence trajectories can be aligned at their peak, as illustrated in the middle panel of \cref{fig:curve_registration}. Once aligned, the trajectories can be averaged to obtain a registered mean curve, which—as shown in the right panel of \cref{fig:curve_registration}—matches the ODE trajectory almost exactly. Here, we aim to apply the reverse process: starting from the deterministic ODE solution, we introduce amplitude scaling and time shifting to generate a collection of trajectories that mimic the variability observed in CTMC simulations. 

\newpage
\section{Additional Figures} \label{appendix:additional_figs}

\begin{figure}[ht]
    \centering
    \includegraphics[width=0.9\linewidth]{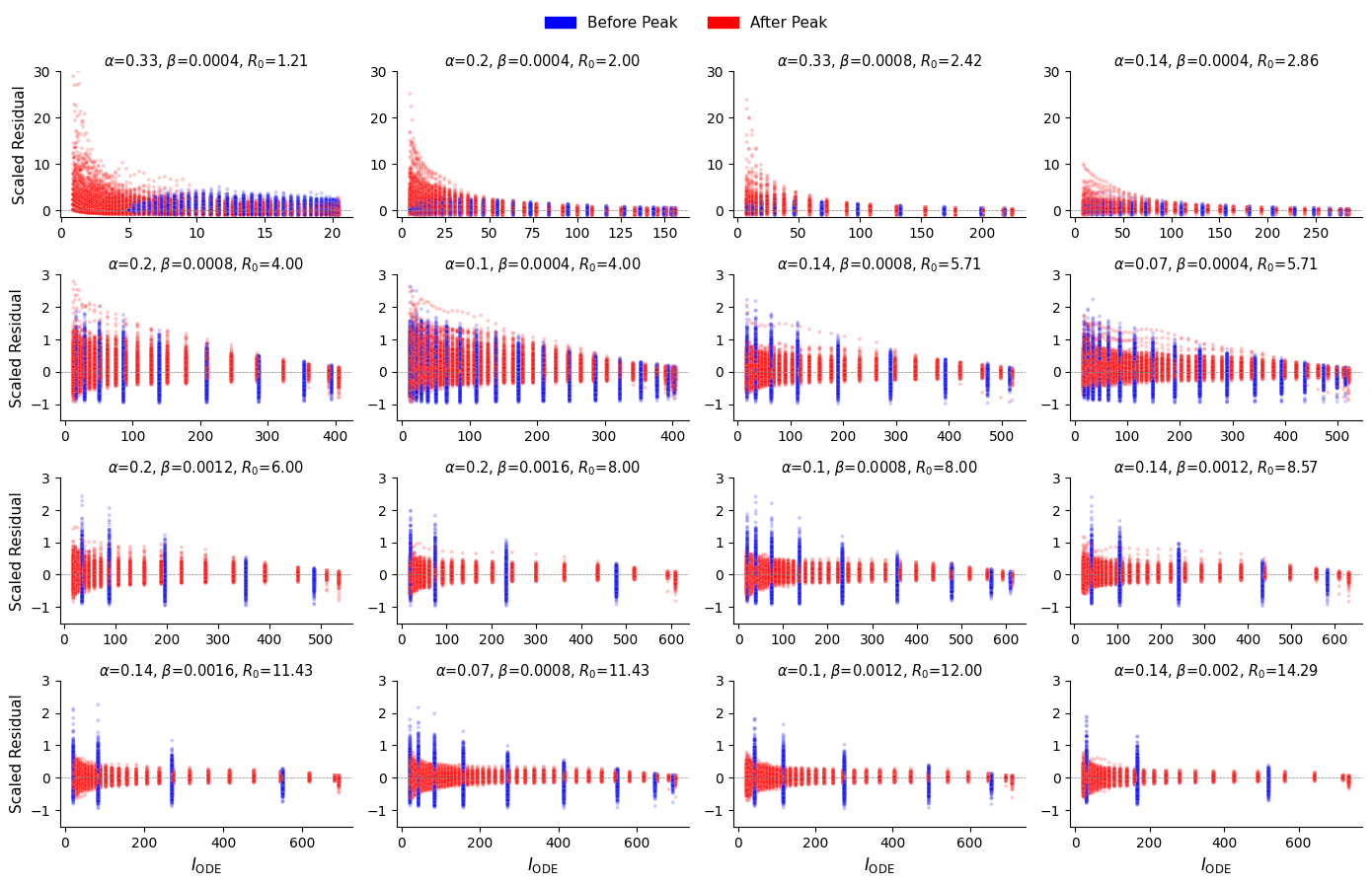}
    \caption{Scaled residuals vs. prevalence counts across all 16 parameter combinations. The top row panels use an extended vertical axis limit to accommodate extreme outliers observed in low $R_0$ settings.}
    \label{fig:placeholder1}
\end{figure}

\begin{figure}
    \centering
    \includegraphics[width=0.9\linewidth]{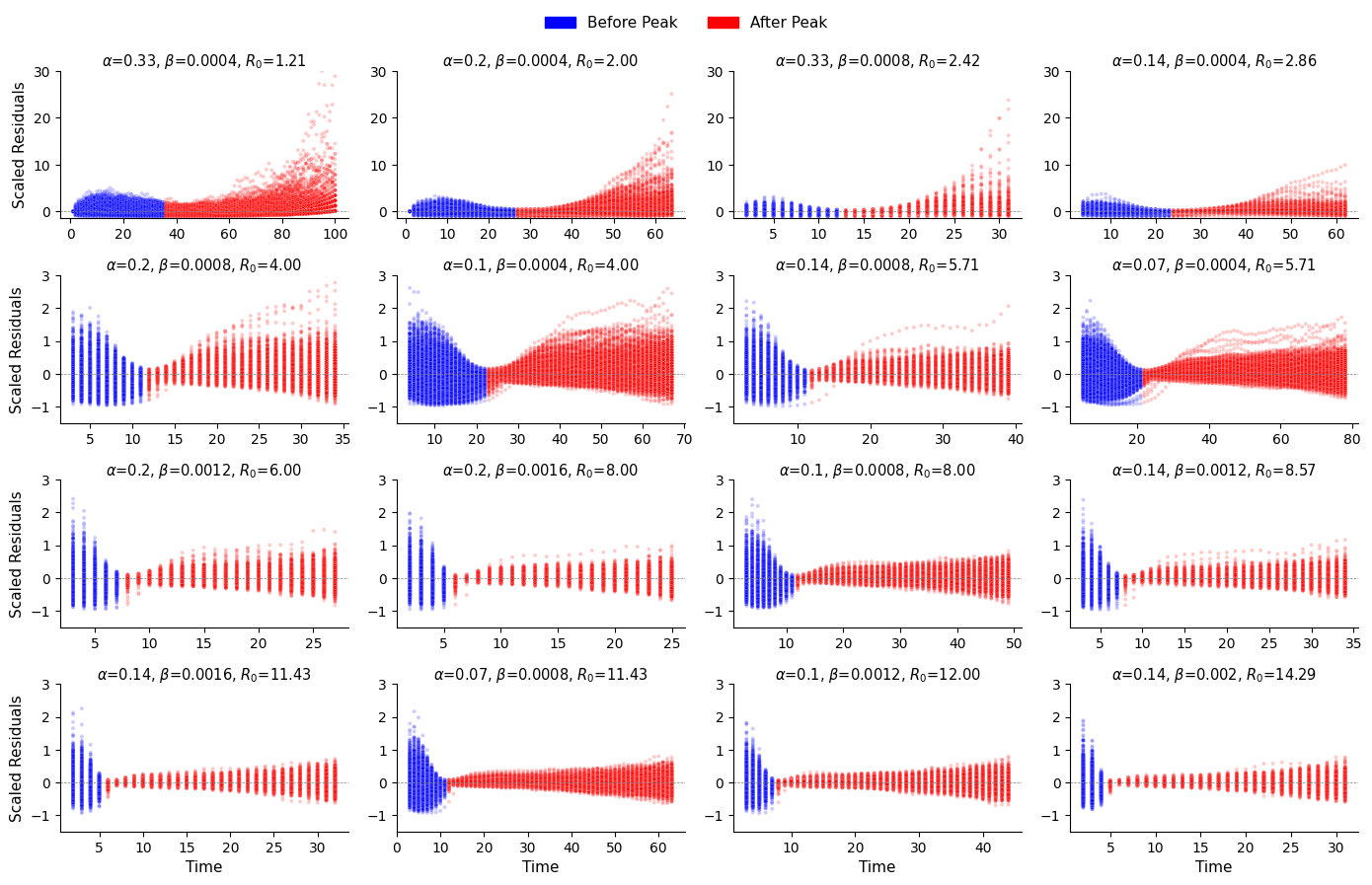}
    \caption{Scaled residuals vs. time across all 16 parameter combinations. The top row panels use an extended vertical axis limit to accommodate extreme outliers observed in low $R_0$ settings.}
    \label{fig:placeholder2}
\end{figure}

\begin{figure}
    \centering
    \includegraphics[width=0.9\linewidth]{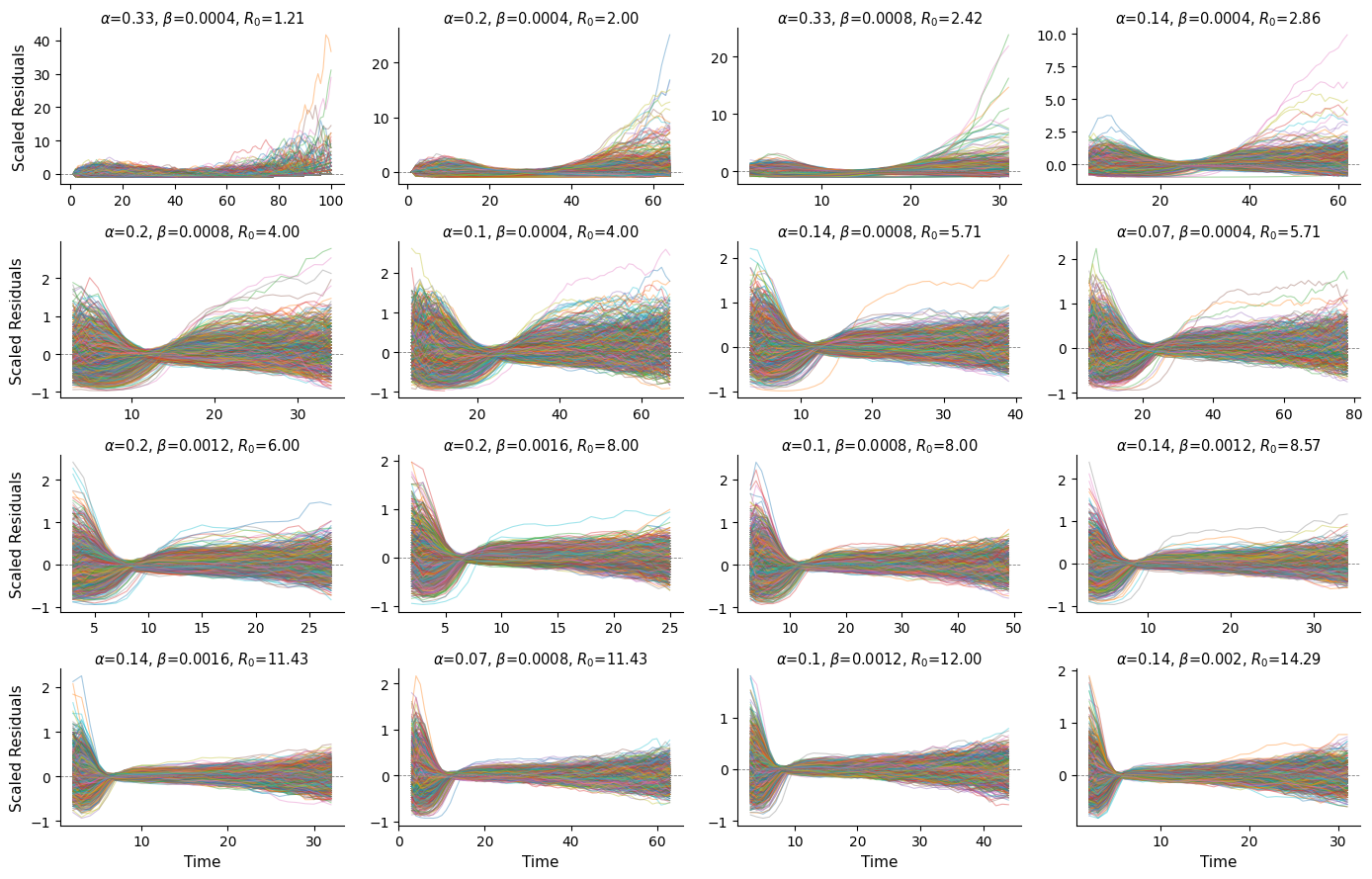}
    \caption{Scaled residual time-series across all 16 parameter combinations.}
    \label{fig:placeholder3}
\end{figure}

\begin{figure}
    \centering
    \includegraphics[width=0.9\linewidth]{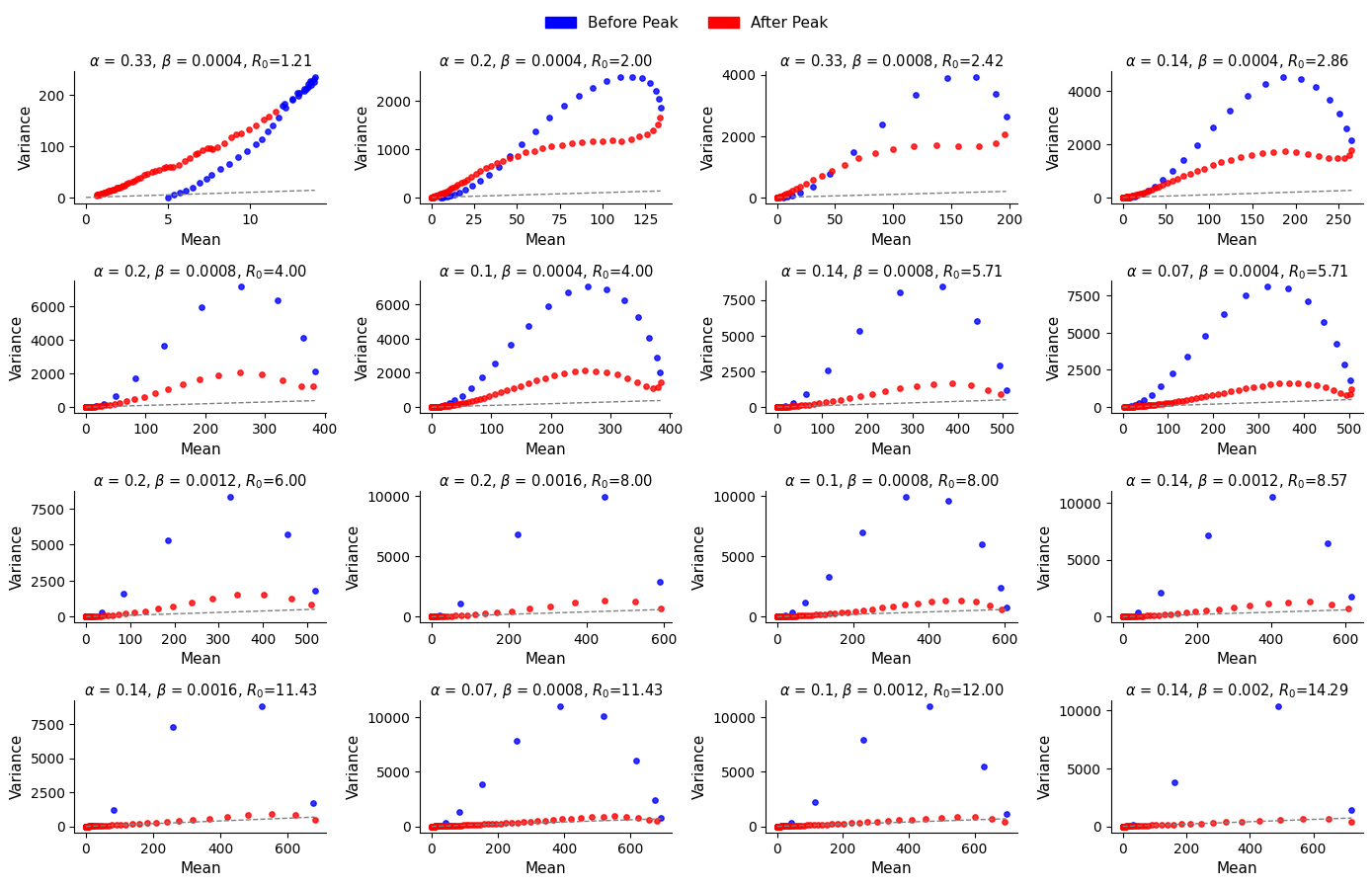}
    \caption{Variance vs. mean of CTMC trajectories for all 16 parameter combinations. Each point corresponds to the variance and mean of the infected population across trajectories at a given time. Blue points denote values before the ODE infection peak and red points denote values after. A dashed gray line (variance = mean) is included in each subplot as a reference for Poisson-distributed behavior, highlighting that all trajectories exhibit super-Poissonian variability.}
    \label{fig:poiss}
\end{figure}

\end{document}